\RequirePackage[hyphens]{url}
\documentclass{sig-alternate}

\usepackage[english]{babel}  
\usepackage{url}

\usepackage{breakurl}
\usepackage[breaklinks]{hyperref}
\usepackage{color}
\usepackage{pifont}

\newcommand{\xmark}{\ding{55}}
\usepackage{newmsc,multicol}

\newcommand{\cs}{cybersecurity} 
\newcommand{\Cs}{Cybersecurity} 
\newcommand{\s}{security} 
\newcommand{\Ss}{Security}

\newcommand{\rev}[1]{\textcolor{black}{#1}}
\makeatletter
\def\@copyrightspace{\relax}
\makeatother
\begin{document}

\doi{10.1108/JIC-05-2019-0127}

\title{Out to Explore the Cybersecurity Planet}

\subtitle{This is a preliminary version of "\href{https://doi.org/10.1108/JIC-05-2019-0127}{10.1108/JIC-05-2019-0127}"
}

\numberofauthors{1} 
\author{
Giampaolo Bella
\\
 \affaddr{Dipartimento di Matematica e Informatica}\\
       \affaddr{Universit\`a di Catania}\\
       \affaddr{Viale A.Doria 6, I-95125, Catania, Italy}\\
       \email{giamp@dmi.unict.it}
}
\date{\today}

\maketitle

\begin{abstract}
\Ss{} ceremonies still fail despite decades of efforts by researchers and practitioners. Attacks are often a cunning amalgam of exploits for technical systems and of forms of human behaviour. For example, this is the case with the recent news headline of a large-scale attack against Electrum Bitcoin wallets, which manages to spread a malicious update of the wallet app.

I therefore set out to look at things through a different lens. I make the (metaphorical) hypothesis that \rev{human ancestors} arrived \rev{on} Earth along with \s{} ceremonies from a very far planet, the \Cs{} planet. My hypothesis continues, in that studying (by huge telescopes) the surface of \Cs{} in combination with the logical projection on that surface of what happens on Earth is beneficial for us earthlings.

I have spotted four cities so far on the remote planet. \emph{Democratic City} features \s{} ceremonies that allow \rev{inhabitants} to follow personal paths of practice and, for example, make errors or be driven by emotions. By contrast, \s{} ceremonies in \emph{Dictatorial City} compel \rev{inhabitants} to comply, \rev{thus behaving} like programmed automata. \Ss{} ceremonies in \emph{Beautiful City} are so beautiful that \rev{inhabitants} just love to follow them precisely. \emph{Invisible City} has \s{} ceremonies that are not perceivable, hence \rev{inhabitants} feel like they never encounter any. Incidentally,  we use the words ``democratic'' and ``dictatorial'' without any political connotation.

A key argument I shall develop is that all cities but Democratic City address the human factor, albeit in different ways.
In the light of these findings, I will also discuss security ceremonies of our planet, such as WhatsApp web login and flight boarding, and explore room for improving them based upon the current understanding of \Cs. 
\end{abstract}

\section{Introduction}
\Cs{} has gone thro\-ugh many theoretical breakthroughs, practical developments, worldwide deployments, subtle flaws and their fixes in a continuous loop --- which should also cover Intellectual Capital (IC) \cite{karenic}. \rev{Often,} \cs{} is \rev{simplistically} understood as a property of a \emph{technical system}, namely one that Scientists design, in the best case along with its \s{} measures, and then pass on for Engineers to build as actual technology. More precisely, that technology consists of interconnected, heterogeneous pieces, such as a browser running on a client host and a server running in the cloud.

\paragraph{The human factor} But \s{} measures continue to fail today. An ever green example comes from the authentication failure due to poor password choice, with the weakest of 5 million passwords leaked in 2018 still being ``123456''  \cite{2018pwd}. A few more examples are outlined below. It is clear that the so called \emph{human factor} may be crucial for the fate of \s{} measures. Therefore, it is insufficient to look at a technical system in all sorts of ways to make sure its \s{} measures work; by contrast, it is necessary to look at the technical system holistically with humans, namely study the effectiveness of the \s{} measures of the \emph{socio-technical system} that intertwines the given technical system with its users.
Therefore, although the mentioned authentication failure is certain to affect the socio-technical system, it may not be entirely due to the inscribed technical system.

Humans may make errors, such as mistakes, namely failures to do what they genuinely wanted to do, or slips, namely momentary lapses that lead to taking an unintended action \cite{norman}. Humans may choose to deliberately counter \cs{} for various reasons, such as perceiving it as burden. For example, although card-and-PIN authentication to enter premises or record work times may be in place, cards are often left in a public card rack outside the entrance near the PIN pad \cite{amazon}. Humans may also fall victims of social engineering scams, hence favour someone else's malicious aims \cite{mitnick}.
We must realise that humans are far from being automata perfectly executing the program that the Technocrats prescribed. 
Hence, scientists \rev{must reinforce their joint} efforts, \rev{particularly with} Technocrats collaborating with colleagues from the Humanities, in order to \rev{improve} their focus \rev{on} a technical system \rev{as just one sub-system of a larger} socio-technical system. As a result, scientists may still pass only a technical system on for Engineers to build, but the resulting technology will be secure when used in practice by humans.

\paragraph{Security ceremonies} Example technical systems are \emph{protocols}, such as the HTTP protocol, and notably protocols that also incorporate \s{} measures, namely \emph{security protocols}, such as the HTTPS security protocol. Correspondingly, example socio-technical systems are \emph{ceremonies}, such as the HTTP ceremony, and notably ceremonies that also incorporate \s{} measures, namely \emph{security ceremonies} \cite{Ellison2007}, such as the HTTPS security ceremony \cite{tls-ceremony}. 

With its emphasis on \s{} and the human factor, this article focuses on \s{} ceremonies. These will be named by their main functional objective, for example a ``flight boarding ceremony''.

\paragraph{Hypothesis (methaphorical)} Security ceremonies are not yet fully understood and are extremely hard to get right, particularly for their inherent human factor. \textbf{The main hypothesis of this article is that \rev{human ancestors} arrived \rev{on} Earth along with \s{} ceremonies from a very far planet, the \Cs{} planet.} Contacts with that planet have been lost entirely, hence humans \emph{and} security ceremonies have been evolving on Earth and on \Cs{} entirely separately.

\rev{The ultimate goal of this article is to understand why \s{} measures continue to fail on our planet and how to strengthen them.
Therefore, the hypothesis of this article also states} that studying (by huge telescopes) the surface of \Cs{} in combination with the logical projection on that surface of what happens on Earth is beneficial for us earthlings \rev{because it gives us a new analysis lens. In particular,} not only will this further our understanding of that planet but, as is the case with any space exploration, it will favour a broader and more structured understanding of what we experience on Earth --- in this case, in terms of security. The results discussed below will establish this hypothesis as fact.

\paragraph{Article contribution} 
This article elaborates on the little we know about the \Cs{} planet. I have spotted four cities so far. \emph{Democratic City} features \s{} ceremonies that allow \rev{inhabitants} to follow personal paths of practice and, for example, make errors or be driven by emotions. By contrast, \s{} ceremonies in \emph{Dictatorial City} compel \rev{inhabitants} to comply, \rev{thus behaving} like programmed automata. Then, \s{} ceremonies in \emph{Beautiful City} are so beautiful that \rev{inhabitants} just love to follow them precisely. Finally, \emph{Invisible City} has \s{} ceremonies that are not perceivable, hence \rev{inhabitants} feel like they never encounter any.

This is, in short, my \rev{current, limited} understanding of \Cs. Following the stated hypothesis, it combines what can be seen of that planet by telescopes \rev{from Earth} with what can be predicted on the basis of terrestrial experience. 
A key argument that I shall develop is that all cities except Democratic City address the human factor, albeit in different ways.
In the light of these findings, I will also discuss security ceremonies of our planet, such as WhatsApp web login and flight boarding, and explore room for improving them upon the basis of the current understanding of \Cs.
 
\paragraph{Lexical considerations}
Because this article is solely about security ceremonies, these will be often referred to simply as `ceremonies'. Similarly, `\s{}' will often be used as a general concept, in fact referring to `\s{} measures'.

Moreover, a  ceremony will be addressed depending on the city in which it is used, so, a ceremony may have the \rev{features of being, respectively} dictatorial, democratic, beautiful or invisible. A ceremony may even combine the four features to some extent, depending on whether it is used between which cities, consequently \rev{getting and variously mixing their respective} influences and customs.

The features are attributes of the \s{} measure that a ceremony uses. 
Therefore, any claim that is made for brevity about a security ceremony should be, more precisely, referred to the part of the ceremony that forms a \s{} measure.
For example, stating that ``a security ceremony is beautiful'' means that its \s{} measure, say an authentication measure, is beautiful.

Humans will be referenced by the male pronoun everywhere, except for the gate attendant, who will be referenced by the female pronoun to avoid ambiguity with the passenger.

\paragraph{Article structure} 
This article continues by describing the current understanding of the four known cities of the \Cs{} planet, precisely Democratic City (\S\ref{demo}), Dictatorial City (\S\ref{dict}), Beautiful City (\S\ref{beauty}) and Invisible City (\S\ref{disappearance}). It develops a critical discussion (\S\ref{disc}) and terminates with some conclusions (\S\ref{concl}).

\section{Democratic City}\label{demo}
Contrarily to the variety of vulnerabilities that have occurred on Earth, such as SSL Heartbleed \cite{heartbleed}, Shellshock \cite{shellshock} and key reinstallation in WPA2 \cite{krack}, vulnerabilities in this city are not only purely technical but leverage the human factor. 
\textbf{In Democratic City,
\s{} measures are hard to interpret, use and comply with for inhabitants. \rev{Inhabitants} may then interact with  \s{} ceremonies differently from what the designers of the ceremonies expected, producing vulnerabilities.} 
For example, some people may opt to abort or disable the automatic software update ceremony. 
Therefore, vulnerabilities exist here despite the Technocrats' efforts at preventing them, and in fact all sorts of security incidents have happened over time.

In general, \s{} in Democratic City resembles a huge number of scenarios noticeable in our planet, where 
the ``IBM Security Services 2014 Cyber Security Intelligence Index'' stated that 
``\emph{over 95 percent of all incidents investigated recognize `human error' as a contributing factor}'' \cite{ibm}. The relevance of the human factor to \s{} has not changed over the last years, with the ``IBM X-Force Threat Intelligence Index 2018'' remarking that ``\emph{the potentially detrimental impact of an inadvertent insider on IT security cannot be overstated}'' \cite{ibm18}.  
\rev{The ``Verizon 2019 Data Breach Investigations Report'' remarks that ``\emph{Errors were causal events in 21\% of breaches}'', that ``\emph{15\% were Misuse by authorized users}'' and that ``\emph{33\% included social attacks}''  \cite{verizon19}.}
To put it simply, humans continue to be somewhat irrational and generally unpredictable despite their risk-benefit mental models \cite{psychoschneier} and the (technical systems') use of motivating values and reward mechanisms \cite{psychowest}.

Democratic City can therefore be understood by projecting the socio-technical attacks that we witness on Earth.
Examples are easy to make. A couple of traditional ones are the sticky notes with login passwords or other security-sensitive information, or the post-completion errors during the log off procedure from a shared or public computer (with the result that the log off does not happen).

But more tricky ones have occurred. One dating back to a few of years ago derived from how a human could share a file he had on Dropbox or on Box with other humans \cite{boxdropbox}. Both systems allowed a human to generate a long, hence difficult to memorise, URL for the file; the human could then send this file-share URL in all sorts of ways to the people he wanted to share the file with. It might have been clear to (almost) everyone that the file-share URL would grant direct access to the file without any form of authentication, hence it was meant to be used with care, nevertheless two subtler vulnerabilities exposing the file were there:
\begin{itemize}
\item
Users inadvertently put the file-share URL in the search bar rather than in the address bar, causing the search engine to index the file-share URL, hence various third parties to access the file.
\item 
If the file was a text file containing a clickable link to a third party address, humans who clicked on that link caused referral data including the file-share URL to be sent to the third party, which was then enabled to access the file.
\end{itemize}

It is clear that these vulnerabilities somewhat get emptied from a purely technical standpoint because the technical systems were running as normal, namely as expected by its designers and without being violated at any rate. By contrast, we are in front of inherently socio-technical vulnerabilities: they stem from a combination of humans' ignorance, misunderstanding, inadvertent behaviour or perhaps even deliberate, stubborn misconception that no vulnerability would affect them. Such stances have variously mixed over time, forming a variety of humans' states of mind or personas that can be considered a large part of the cause of the vulnerabilities.

Along these lines, even the established policy of asking humans to change their passwords from time to time may falter. It is sound in purely technical terms because every secret could be discovered with infinite resources. But a socio-technical reading of it reveals the vulnerabilities tied to the reiterated human choices of passwords. These are customarily thwarted by technical checks requiring the password to be strong enough. But it was recently found out that humans often resort to simple, algorithmic changes of the previous password to build the new one, hence attackers just have to fine-tune their brute-forcing techniques \cite{cranor}.

We also know that \s{} often intertwines with people's safety. The 2015 ``Chatham House Report'' shouts out loud that ``\emph{Some 
nuclear facilities do not change the default passwords on their equipment}'' \cite{chatham}, which implies that attackers may sneak in by trying out default passwords such as `0000' or `1234'. It is hard to believe that highly trained and skilled operators do not change the default passwords (because of inadvertence, laziness or what sort of distorted logic), but this is reported to be the case even if public safety is involved.

It can be conjectured that these observations also apply to Democratic City on the \Cs{} planet. It then seems fair to conclude that humans are somewhat \emph{incorrectly} dealt with in this city because a large part of the responsibilities for many \s{} failures seem to fall on them. \rev{In other words, possible threats due to human interaction are not well accounted for}. The question then arises as to whether there exist other cities on the planet where humans integrate with the technical systems more smoothly, namely in such a way that the resulting ceremonies are secure. The sequel of this manuscript shows that the answer could be affirmative.

\section{Dictatorial City}\label{dict}

I spotted another city on the \Cs{} planet, Dictatorial City.
Here, humans interact with the technical systems in a nearly opposite fashion to that of Democratic City. \textbf{\rev{Inhabitants}' freedom and responsibility of making choices is reduced to a minimum, such as to determining whether to initiate a ceremony 
or not}.
If they opt to engage, then the ceremonies are fully automated and compel humans to specific interactions, somewhat in the style of Poka-yoke \cite{pokayoke}. Therefore, humans are not left the choice of deciding how to take a specific step with the system and what information or object or device to use in that step. This does not mean that humans

have lost their potential maliciousness or are infallible, but merely that the ceremony receives them securely, namely without hindering its security measures.

So, in this city, human interactions with the technical systems are fully determined by the latter. The only reliance on humans is to memorise simple (na\-me\-ly short and trivial) secrets or carry predetermined objects such as identity documents, or special devices that could be some evolution of RFID cards or smart tokens. Therefore, humans are never required to invent and retain by heart strong (na\-me\-ly long and complex) secrets, and the Dropbox/Box vulnerabilities mentioned above do not exist here because the corresponding features are inhibited by the technical systems.
In other words, the ceremonies enforce limited interactions with humans, hence those interactions cannot overturn the \s{} measures.

We do not know much more, yet, about Dictatorial City, but can try again to understand it by projection of what we encounter on Earth. It can be expected that humans are never required to make those difficult and often casual choices on whether and how to continue in front of a web site whose certificate the browser cannot validate, because a malicious man-in-the-middle cannot be ruled out \cite{jbrowser}.  
Similarly, this city does not host an SSH configuration that is popular on Earth.
When a human connects to a remote host via SSH for the first time, the system displays a message of the form:
\begin{verbatim}
  The authenticity of host www.dmi.unict.it can't
  be established. 
  RSA key fingerprint is 
  2b:05:ff:64:91:60:24:3a:6e:83:c7:7a:c5:85:0a:41. 
  Are you sure you want to continue connecting
  (yes/no)?
\end{verbatim}
By denouncing the lack of a viable public-key certification system, the human is asked a very nasty question that will normally get a blind `yes' answer because out-of-band verification of the fingerprint is tedious. The risk of man-in-the-middle rises again, especially because SSH could be used from a remote, untrusted location such as a cafe. (If the human continues the first time, the fingerprint is then stored locally and the message will no longer be displayed, hence a successful attack would become persistent and stealth). 

By contrast, a \s{} ceremony that can exist in this city
could be inspired to the recent NIST guidelines on authentication, which contain what can be termed the ``NIST 2017 Password Setup'' ceremony \cite{nistnewreport}. It rests on the security measure of \rev{having the computer choose each secret pseudorandomly by executing an algorithm and also set a threshold for failed verification attempts (to thwart password bruteforcing)}. Then, secrets can be as short as ``\emph{6 characters in length and MAY be entirely numeric}''  \cite{nistnewreport}. This subverts the previous and currently most widespread ``NIST 2004 Password Setup'' ceremony, \rev{whose implementation} constrained the human to invent a robust password (of sufficient length and with alphanumeric and special characters) \cite{nist}. The newer ceremony is the more dictatorial because it fully removes the human choice of a password. The older ceremony is, in turn, more dictatorial than the pre-existing ceremony, say ``Unconstrained Password Setup'', which allowed the human to even choose a weak password.

It is not fully understood how the technical systems are administered in Dictatorial City. We do not know whether such systems are intelligent enough to maintain themselves without human intervention but conjecture that such intervention, if needed, is not malicious and is correct --- as if at least humans who are administrators were well-programmed automata. Both options would take to a logical extreme, for example, the ways mandatory access control and military systems are administered on Earth.

In summary, human interaction will never undermine \s{} ceremonies in this city of the \Cs{} planet. Unfortunately, we do not know how to fully accomplish this on Earth as, for example, existing measures to prevent humans from deliberately sharing their secrets are not fully effective and scalable. But neither Dictatorial City is ideal, as I have spotted no evidence that \s{} ceremonies here do not suffer purely technical issues. In other words, there may still exist technical bugs, hence not due to
interactions with humans, thus in the style of the examples made above \cite{heartbleed,shellshock,krack}.

\section{Beautiful City}\label{beauty}
\textbf{Security is beautiful in this city, hence \rev{inhabitants} well receive the \s{} ceremonies, are attrac\-ted to them and  naturally use them as prescribed by their designers}. A logical consequence of this assumption is similar to what we observed in Dictatorial City (\S\ref{dict}), that whatever vulnerabilities a \s{} ceremony might have would not be due to how humans execute the ceremony, but to technical bugs. However, humans here are not somewhat enslaved but, rather, fulfilled by the interaction with the technical systems.

Inspired by as little as we know about this city, I have been working on trying to reproduce a similar human experience of \s{} on our planet.
Vigan\`o and I postulated that \s{} is beautiful if it satisfies a triple of abstract requirements: to be a primary system feature, not to be disjoint from the system functions to be secured, and to be ambassador of a positive user experience  \cite{spw15}. I am going to expand them below.

The first one is unsurprising by itself, as it appeals to the security-by-design principle that we have known for years. Therefore, a technical system, say a web site, should be designed with security in mind since the beginning; this normally enhances the effectiveness of the  \s{} measures of the system without making it overly clumsy. Experts in \s{} should then engage with and contribute to the design of the technical system since the design inception. This is also the case with kitchen appliances, for example, because these are getting more and more interconnected through the IoT. For example, when security is not fully accounted for by design in this area, the crooks could be granted a ``\emph{potential way to steal humans' Gmail credentials from a Samsung smart fridge}'' \cite{fridge}. The fridge failed to validate SSL certificates, thus enabling a man-in-the-middle attack.

The second requirement insists on what even security-by-design fails to prescribe very clearly, that the secure access to a system be exclusive, namely the only possible one. In other words, there is no point building a strong security ceremony without disabling a simple bypass. The relevance of this requirement can be appreciated by evaluating (a variety of) scenarios that have neglected it on Earth. For example, let us think of a web site secured via HTTPS, yet allowing access also via HTTP for whatever legacy or performance reasons; the latter obviously puts credentials at risk. 

It is worth mentioning yet another attack that becomes possible if the second requirement is neglected. It sees a human access a Wi-Fi network using whatever security protocol that does not prescribe SSID validation: it becomes trivial for an attacker to setup a fake SSID with the same name as the target SSID and harvest clients' login credentials upon the odds that the clients chose the fake SSID. 

The third requirement of beautiful security is perhaps the most abstract one yet most crucial. \Ss{} ought to be positive, nice, rewarding and, generally speaking, a somewhat desirable thing to have and comply with. While the bottom line of this feature may be subjective, the requirement aims at something that \emph{can be generally considered} positive. (For example, despite subjectivity, it might be hard to find people who would genuinely dislike Nutella spread or Ferrari cars).

There is at least another relevant example of something very popular on Earth whose perception has been fully upturned from negative into positive: the use of chewing gum. When I went to primary school, chewing gum was forbidden because ``\emph{it causes cavity!}'' but quite the opposite holds today. Thanks to simple changes in the ingredients, such as removing sugar and adding fluorine, chewing gums are often used to thwart cavity when tooth brushing is not handy. I advocate a similar twist of the plot to happen with \s{} ceremonies through beautiful \s{}.

A possible twist could occur by seeking out to design
ceremonies as an engaging and fun game \cite{hendrix2016game}. An episode of the Peppa Pig cartoon portrays a group of kids wanting to be part of a ``\emph{secret club}'' as soon as they come to know of its existence, and they are very willing to pronounce a password \cite{peppa}. Where people's perception of \s{} is negative, can we manage to upturn it into such a positive one as is in the cartoon?
For example, inspired by existing work \cite{gamauth}, we could think of a ``Gamified Password Setup'' ceremony whereby a human derives his 
password through a game that is certain to yield a strong secret.

\begin{figure*}
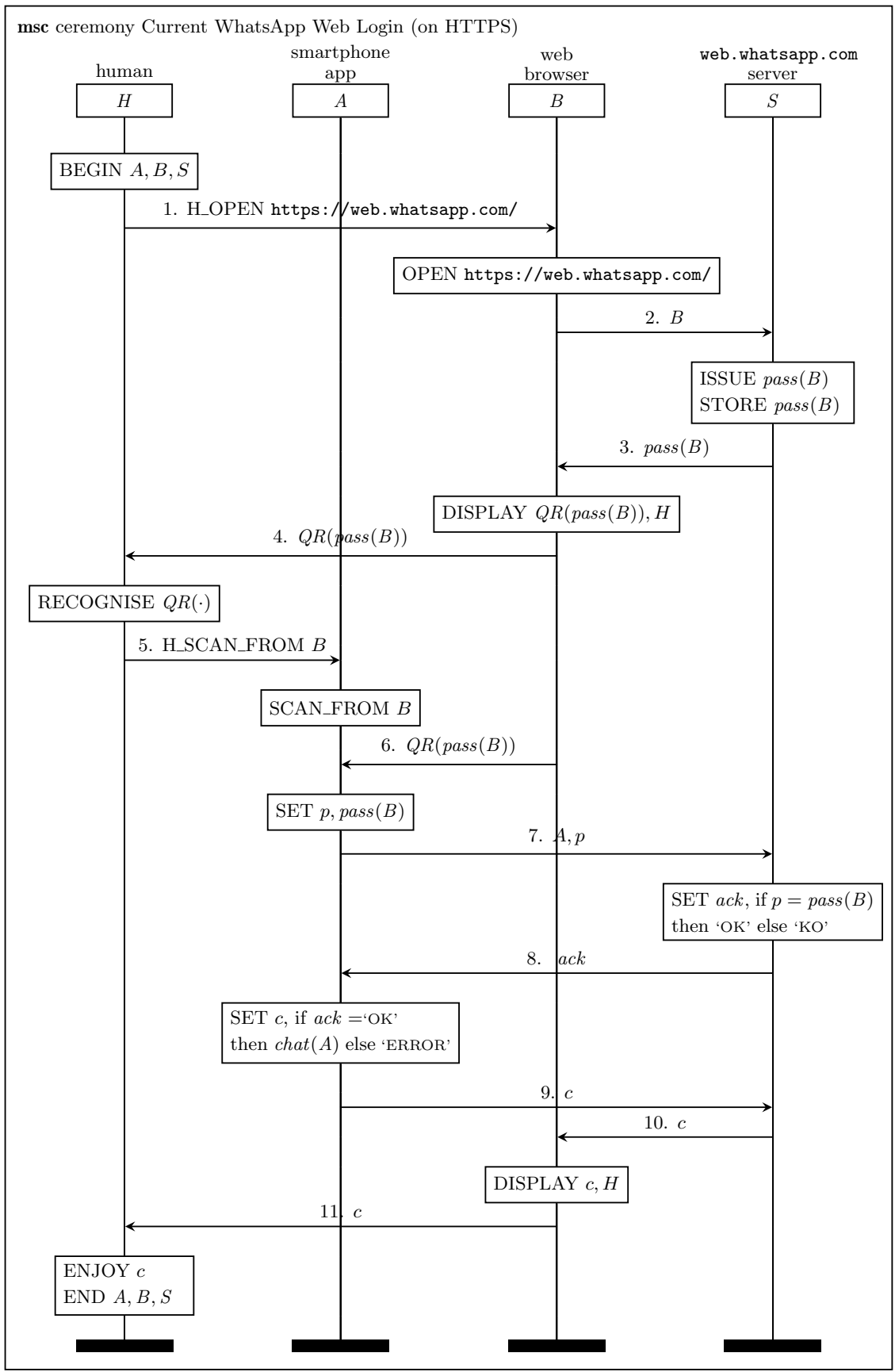

\begin{center}
\begin{msc}{ceremony Current WhatsApp Web Login (on HTTPS)}
\setlength{\instdist}{2cm}
\setlength{\envinstdist}{2cm}
\setlength{\bottomfootdist}{.3cm}
\declinst{h}{human}{$\mathit{H}$}
\declinst{a}{\parbox{2.45cm}{\centering smartphone\\[-.2ex] app\\[2.2ex]}}{$\mathit{A}$}
\declinst{b}{\parbox{2.45cm}{\centering web\\[-.2ex] browser\\[2.2ex]}}{$\mathit{B}$}
\declinst{s}{\parbox{2.45cm}{\centering {\tt web.whatsapp.com}\\[-.2ex] server\\[2.2ex]}}{$\mathit{S}$}
\action*{BEGIN $A,B,S$}{h}
\nextlevel[2.5]
\mess{1. H\_OPEN {\tt https://web.whatsapp.com/}}{h}{b}
\nextlevel
\action*{OPEN {\tt https://web.whatsapp.com/} }{b}
\nextlevel[2.5]
\mess{2. $\mathit{B}$}{b}{s}
\nextlevel
\action*{\parbox[l]{2.45cm}{ISSUE $\mathit{pass(B)}$\\[.5ex] STORE $\mathit{pass(B)}$}}{s}
\nextlevel[3.5]
\mess{3. $\mathit{pass(B)}$}{s}{b}
\nextlevel
\action*{DISPLAY $\mathit{QR(pass(B))}, H$}{b}
\nextlevel[2]
\mess{4. $\mathit{QR(pass(B))}$}{b}{h}
\nextlevel
\action*{RECOGNISE $\mathit{QR(\cdot)}$}{h}
\nextlevel[2.5]
\mess{5. H\_SCAN\_FROM $B$}{h}{a}
\nextlevel[1]
\action*{SCAN\_FROM $B$}{a}
\nextlevel[2.5]
\mess{6. $\mathit{QR(pass(B))}$}{b}{a}
\nextlevel
\action*{SET $p,\mathit{pass(B)}$}{a}
\nextlevel[2]
\mess{7. $A,p$}{a}{s}
\nextlevel[1]
\action*{\parbox[l]{3.4cm}{SET $\mathit{ack},$ if $p=\mathit{pass(B)}$\\[.5ex] then {\scriptsize `OK'} else {\scriptsize `KO'}}
}{s}
\nextlevel[3]
\mess{8. $\mathit{\,\,ack}$}{s}{a}
\nextlevel
\action*{\parbox[l]{3.7cm}{SET $c,$ if $\mathit{ack}=${\scriptsize `OK'}\\[.5ex] then $\mathit{chat(A)}$ else {\scriptsize `ERROR'}}
}{a}
\nextlevel[3.5]
\mess{9. $c$}{a}{s}
\nextlevel
\mess{10. $c$}{s}{b}
\nextlevel
\action*{DISPLAY $c, H$}{b}
\nextlevel[2]
\mess{11. $c$}{b}{h}
\nextlevel
\action*{\parbox[l]{2.05cm}{ENJOY $c$\\[.5ex] END $\mathit{A,B,S}$}
}{h}
\nextlevel[2]
\end{msc}
\end{center}
\caption{The Current WhatsApp Web Login ceremony}\label{WhatsApp}
\end{figure*}

I conjecture that the use of the web interface of WhatsApp on our planet conforms to the beautiful security principle. Figure \ref{WhatsApp} shows my understanding of the ``Current WhatsApp Web Login'' ceremony, featuring the human, the app running on the human's smartphone, the browser running on the human's computer, and the WhatsApp server. 
Activities are written in capital letters. They may be conducted locally, namely appear inside a square box such as DISPLAY for the browser's activity of portraying some data, or ENJOY for the human's full use of the chat; alternatively, they may be \emph{dictated} by the human, hence they (become meta-activities and) replace the information that annotates the arrows, such as H\_OPEN for the human's activity of instructing a browser to open a server.
Steps 1, 4, 5 and 11 involve human activities; in particular, 4 and 11 take place over visual channels with the human \cite{channels}.
The other steps form the technical part of the ceremony; in particular, step 6 occurs over an optical channel, as it consists in the activity of scanning through a camera, while all other technical steps run through TLS-protected, hence authenticated and confidential, channels.

It can be seen that
the browser transmits its identifier to the server, which issues a passcode for the browser, stores it and sends it across. The browser displays it as a QR code, namely sends it to the human on a visual channel. The human recognises it as \emph{some} QR code (hence the notation hides the parameter) and operates the app running on his smartphone to scan it.
The app sends the passcode (just read through the QR code) to the server, which matches it to its stored version and sends the corresponding \rev{acknowledgment} message. Only if the \rev{acknowledgment} is positive does the app output the chat to the browser. The browser finally makes the chat available to the human through the visual channel.

The main goal of the ceremony is to authenticate the browser to the app so that the latter can securely share its chat with the former in step 9. This is coherent with the official WhatsApp policy, which claims that the server does not store any chats. Although this version of the ceremony makes it evident that the server facilitates the browser-to-app authentication, the full ceremony details are proprietary. For example, it is not fully clear how exposing the chat on the server preserves end-to-end encryption.

In terms of beautiful security, I find it most important to stress that the passcode is 128 characters long, hence it would have been too tedious for the human to read from the browser and tap in the app, and a ``Tap-in WhatsApp Web Login'' ceremony would have been hardly beautiful. Having the human point the phone to the browser and scan the QR code is (a crucial design choice that is) being well received --- the vast use of the web interface of WhatsApp supports the claim that QR-code scanning conjugates usability, simplicity, security and also some beauty. 

This example gives me hope that beautiful security can be reached also on Earth. However, considerable effort will be needed, for example involving large-scale human studies to distill out the features of beauty that \s{} ceremonies could leverage in general.

\section{Invisible City}\label{disappearance}

\textbf{In Invisible City, \rev{inhabitants} cannot see the \s{} measures of ceremonies, \rev{hence}, \s{} is not perceivable by humans although it is still there}. Humans are able to conduct the somewhat obvious activities to pursue their intended goals while not worrying about their security implications. Those activities are secure anyway, yet without any apparent \s{} measures in place. Therefore, I am trying to spot, for example, whether in this city a human can access his bank account \emph{securely} on any electronic terminal he merely \rev{stares} at, coherently with what we know, that ``\emph{The ideal security user experience for most users would be none at all}'' \cite{psychowest}. In parallel with seeking such evidence,  
I have also been studying how to make \s{} ceremonies more invisible on Earth.
Vigan\`o and I suggested integrating \s{} measures with functions or, alternatively, with other measures that humans would already accept as routine \cite{spw16}. A similar notion of invisibility has been recently suggested in the context of system patching \cite{disappearing}. It is a reversal of the defense-in-depth principle for the sake of improving the user experience, yet preserving the overall security.

A pioneering example is the Iphone 5S's integration of the screen wake-up button with the fingerprint sensor. This idea stemmed from the observation that people were used to a stand-by display being off to preserve battery, hence to the need of pressing some button to wake it up when they sought to use it. Integration in this case combines a routine activity, such as screen wake-up, with an important security measure, namely user authentication to the phone, and here is the resulting ``iPhone 5S Wake-up'' ceremony. By contrast, the previous ``iPhone 5 Wake-up'' ceremony was based on pressure of the wake-up button and would continue with the separate and more traditional authentication measure of tapping a PIN in. 

Other examples of how to make \s{} ceremonies more invisible than before can be drawn from the integration of two \s{} measures. One is the use of one password to both decrypt a hard disc and access a user account; another one is the use of a single button on cars' remote controls to both toggle the power-door locks and the alarm system. However, two separate measures could be harder to violate in certain scenarios.

\begin{figure*}[h!]
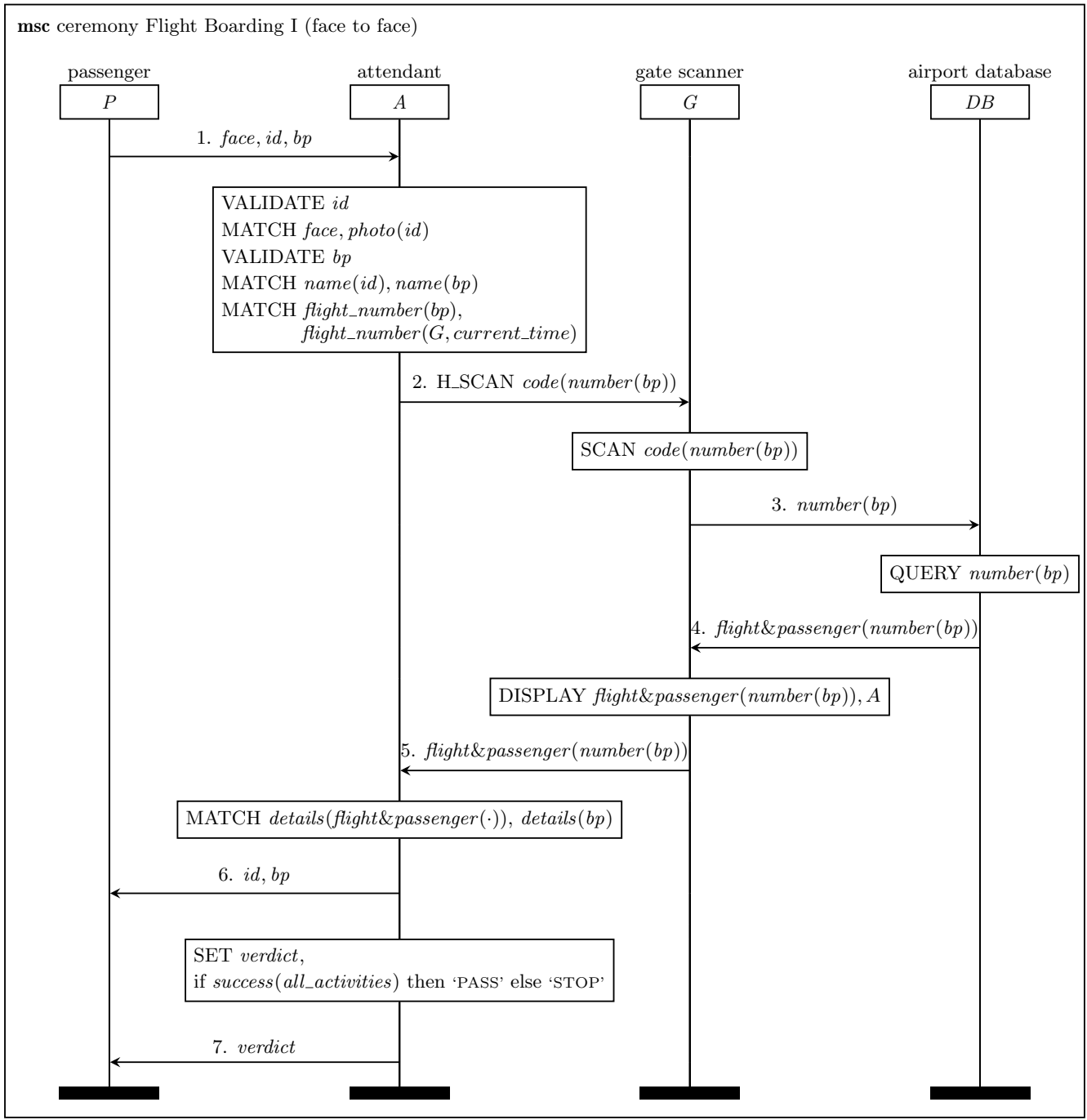

\begin{center}
\begin{msc}{ceremony Flight Boarding I (face to face)}
\setlength{\instdist}{31mm}
\setlength{\envinstdist}{17mm}
\setlength{\bottomfootdist}{.3cm}
\declinst{p}{passenger}{$\mathit{P}$}
\declinst{a}{attendant}{$\mathit{A}$}
\declinst{g}{gate scanner}{$\mathit{G}$}
\declinst{d}{airport database}{$\mathit{DB}$}
\mess{1. $\mathit{face,id,bp}$}{p}{a}
\nextlevel
\action*{\parbox[l]{5.8cm}{VALIDATE $\mathit{id}$\\[.5ex] MATCH $\mathit{face,photo(id)}$\\[.5ex] VALIDATE $\mathit{bp}$\\[.5ex] MATCH $\mathit{name(id),name(bp)}$\\[.5ex]MATCH $\mathit{flight\_number}(bp)$,\\ \hspace*{12,2mm} $\mathit{flight\_number}(G, current\_time)$}}{a}
\nextlevel[7]
\mess{2. H\_SCAN $\mathit{code(number(bp))}$}{a}{g}
\nextlevel[1]
\action*{SCAN $\mathit{code(number(bp))}$}{g}
\nextlevel[3]
\mess{3. $\mathit{number(bp)}$}{g}{d}
\nextlevel
\action*{QUERY $\mathit{number(bp)}$}{d}
\nextlevel[3]
\mess{4. $\mathit{flight\& passenger(number(bp))}$}{d}{g}
\nextlevel
\action*{DISPLAY $\mathit{flight\& passenger(number(bp))}, A$}{g}
\nextlevel[3]
\mess{5. $\mathit{flight\& passenger(number(bp))}$}{g}{a}
\nextlevel
\action*{MATCH $\mathit{details(flight\& passenger(\cdot))}$, $\mathit{details(bp)}$}{a}
\nextlevel[3]
\mess{6. $\mathit{id,bp}$}{a}{p}
\nextlevel[1.5]
\action*{\parbox[l]{6.7cm}{SET $\mathit{verdict},$\\[.5ex]if $\mathit{success(all\_activities)}$ then {\scriptsize `PASS'} else {\scriptsize `STOP'}}}{a}
\nextlevel[4]
\mess{7. $\mathit{verdict}$}{a}{p}
\end{msc}
\end{center}
\caption{The traditional flight boarding ceremony with attendant's full checks}\label{boardingI}
\end{figure*}

Airport security offers a remarkable example here. It has become extremely relevant at least because it comes to support passengers' safety. It derives from a variety of possibly interconnected security ceremonies, the main ones aimed at passengers' check-in, security controls, and flight boarding. Let us focus on the last one, which takes place face-to-face between a passenger and a gate attendant, with the participation of the gate scanner and a database as technical systems. My understanding of the traditional, most widespread version is shown in Figure \ref{boardingI}, ceremony ``Flight Boarding I''. It is the version that sees the attendant perform the full checks to authenticate the passenger and determine that he is authorised to board the particular aircraft waiting behind the gate.
The ceremony is presented with the same style used above for activities and channels. In particular, a meta-activity also appears here, H\_SCAN, which also delivers a (bar or QR) code
to the gate scanner, thus it occurs on an optical channel. Steps 3 and 4 are technical and assumed secure. The remaining steps unfold over visual channels, in particular step 5 on a channel between attendant and gate scanner, and steps 1, 6 and 7 on a channel between passenger and attendant.

Passenger $P$ gives gate attendant $A$ three pieces of information: the passenger's face, his id, such as an identity card or a passport, and his boarding pass. Traditionally, the boarding card would be on a special paper, but has lately evolved to a printout or to a version displayed through the passenger's smartphone.
The attendant begins by checking id authenticity and validity (e.g. it has not expired), and then matching face to id photo. The attendant then checks validity of the boarding pass (e.g. its date is correct) and matches the id name to the boarding pass name. Another important match is between the flight number reported on the boarding pass and the flight number currently assigned to that gate. When all these five activities succeed, the attendant has authenticated the passenger and has some evidence, through the boarding pass, that he is authorised to board the particular aircraft at the gate.
Up to approximately the mid 1990s (depending on airports), the ceremony would jump straight onto step 6 (of Figure \ref{boardingI}) because a valid boarding pass was considered sufficient authorisation evidence. In fact, the use of database technologies allows the attendant to seek extra authorisation confirmation from the airport database. The attendant then scans the (bar or QR) code of the boarding pass number through the gate scanner, and this causes a query to the database, whose outcome is displayed. The attendant then matches the details of flight and passenger as displayed with those on the boarding pass (the notation hides the parameter of the former because the attendant trusts the scanner to display details about the boarding pass just scanned, hence the attendant does not need to find the boarding pass number through those details).
Only if this activity succeeds too, is the passenger allowed to go through, otherwise he is stopped for further scrutiny. 

I observe that there are a number of activities for the attendant to carry out per each passenger of a long boarding queue, precisely two VALIDATE and four MATCH activities. These may turn into a source of tiredness or boredom, hence cause errors or deliberate activity deviations or simplifications. In support of this conjecture comes a headline that saw a passenger complain to have taken a wrong flight, reaching a wrong destination \cite{Mirror}. Another similar event occurred recently \cite{ryanairpisa}. Arguably, the attendant got at least her final MATCH activity (of Figure \ref{boardingI}) somewhat wrong. With airport security being so sensitive at present, this event could be turned into various threats should a passenger attempt it deliberately and not report it. Threats may range from terrorists targeting a specific flight without disclosing its purchase, to business threats such as a passenger hopping on a more expensive flight than the purchased one, or even sending someone else to fly on his ticket without paying the fee. These could become more effective by exercising some social engineering activities on the attendant, such as simple distraction.

I remark that this is my own version of a widespread boarding ceremony and, of course, there are many variants also in use. One rests on an electronic boarding pass shown on the passenger's smartphone through a dedicated app or as a PDF document. In my personal experience, it would seem that, with this ceremony variant in use, the attendant routinely dismisses the ``VALIDATE $\mathit{bp}$'' activity, plausibly because it is not clear how to do that. However, an electronic version of a boarding pass is at least equally easy to fake as the paper version. This choice of the attendant's signifies that she is opting to offload the verification effort on the technical systems, because she is assuming that the gate scanner can take that effort and check the pass against a relevant database. A sibling observation is that there is some redundancy in the attendant's activities aimed at checking passenger authorisation, as these consider both information she gathers from the boarding pass and information she reads from the display integrated with (or connected to) the scanner.

\begin{figure*}[h!]
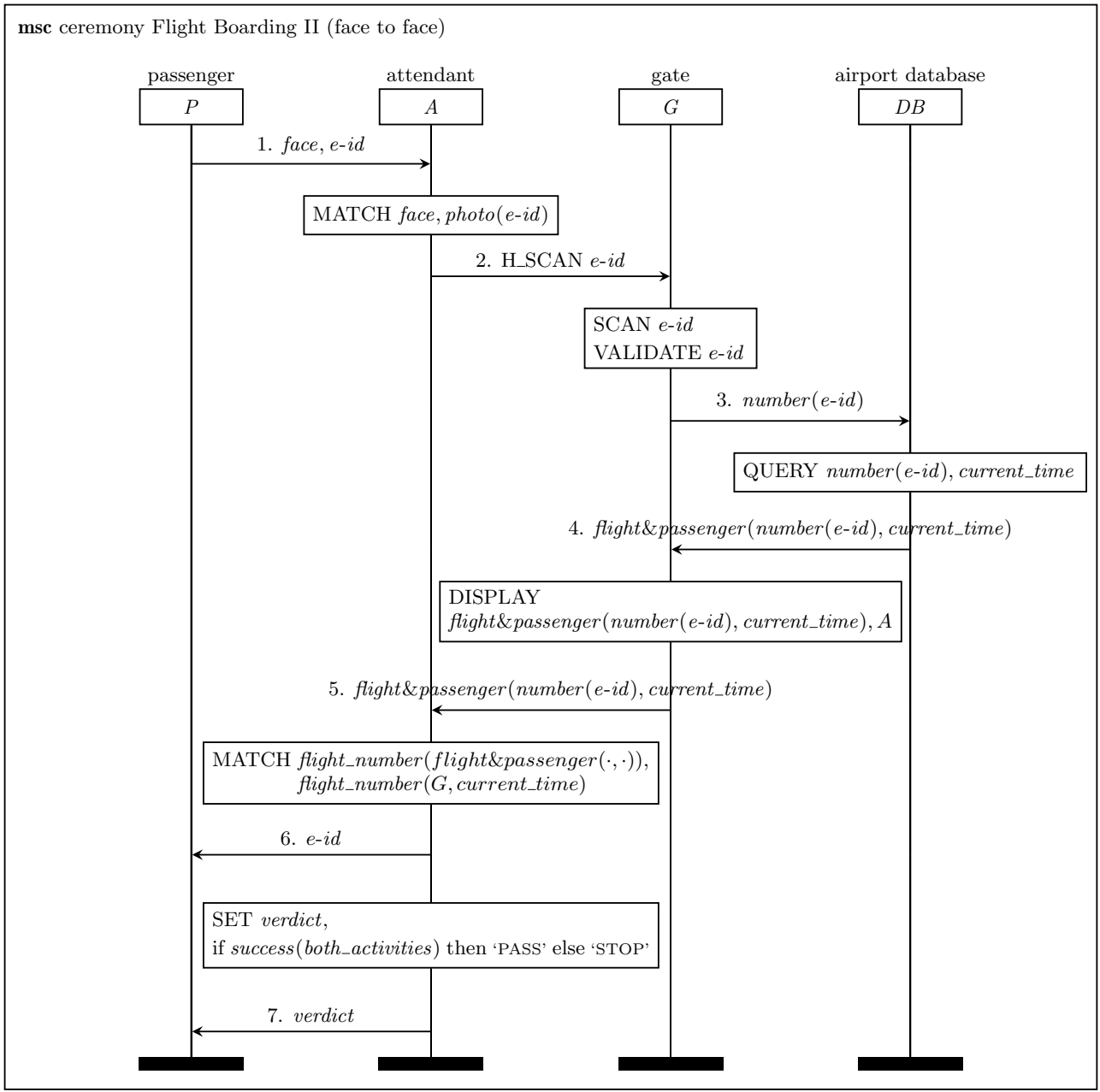

\begin{center}
\begin{msc}{ceremony Flight Boarding II (face to face)}
\setlength{\instdist}{21mm}
\setlength{\envinstdist}{29mm}
\setlength{\bottomfootdist}{.3cm}
\declinst{p}{passenger}{$\mathit{P}$}
\declinst{a}{attendant}{$\mathit{A}$}
\declinst{g}{gate}{$\mathit{G}$}
\declinst{d}{airport database}{$\mathit{DB}$}
\mess{1. $\mathit{face,e}$-$\mathit{id}$}{p}{a}
\nextlevel
\action*{MATCH $\mathit{face,photo(e}$-$\mathit{id)}$}{a}
\nextlevel[2.5]
\mess{2. H\_SCAN $e$-$\mathit{id}$}{a}{g}
\nextlevel[1]
\action*{\parbox[l]{24mm}{SCAN $e$-$\mathit{id}$\\[.5ex] VALIDATE $e$-$\mathit{id}$}}{g}
\nextlevel[3.5]
\mess{3. $\mathit{number(e}$-$\mathit{id})$}{g}{d}
\nextlevel
\action*{\parbox[l]{5.2cm}{QUERY $\mathit{number(e}$-$\mathit{id), current\_time}$}}{d}
\nextlevel[3]
\mess{4. $\mathit{flight\& passenger(number(e}$-$\mathit{id), current\_time})$}{d}{g}
\nextlevel
\action*{\parbox[l]{6.9cm}{DISPLAY\\ $\mathit{flight\& passenger(number(e}$-$\mathit{id), current\_time}), A$}}{g}
\nextlevel[4]
\mess{5. $\mathit{flight\& passenger(number(e}$-$\mathit{id), current\_time})$}{g}{a}
\nextlevel
\action*{\parbox[l]{68mm}{MATCH $\mathit{flight\_number}(flight\& passenger(\cdot,\cdot))$,\\ \hspace*{12,2mm} $\mathit{flight\_number}(G, current\_time)$}}{a}
\nextlevel[3.5]
\mess{6. $e$-$\mathit{id}$}{a}{p}
\nextlevel[1.5]
\action*{\parbox[l]{6.8cm}{SET $\mathit{verdict},$\\[.5ex]if $\mathit{success(both\_activities)}$ then {\scriptsize `PASS'} else {\scriptsize `STOP'}}}{a}
\nextlevel[4]
\mess{7. $\mathit{verdict}$}{a}{p}
\end{msc}
\end{center}
\caption{A flight boarding ceremony where security is more invisible than in Flight Boarding I}\label{boardingII}
\end{figure*}

These notes inspire a process to make the security measures (of passenger authentication and authorisation) of this ceremony more invisible for its human participants, for the attendant especially. 
One way would be to offload the measures as much as possible on the technical components, provided these exist. I therefore suggest to completely dispose with the boarding pass and leverage an electronic id, which can be easily scanned. The resulting ceremony, ``Flight Boarding II'', is given in Figure \ref{boardingII}. Its security is certain to have become more invisible for the passenger because he does not need to carry a boarding pass anymore. It is substantially more invisible for the attendant, who is merely left with two MATCH activities altogether. 
Two of the early activities of the attendant's are no longer necessary, and two are postponed. In particular, id validation is postponed but demanded to the gate, which is therefore no longer a simple scanner (it will perform some cryptographic verification of some digital certificate). The verification that the passenger is at the right gate is still for the attendant to carry out, though deferred till after step 5, when the relevant information becomes available. It is a match between the flight number currently assigned to the gate, which is public information, and the flight number as it is displayed to the attendant following the boarding pass scan.

A technical observation is that the database is queried by the electronic id number with the addition of the current time in order to retrieve the most imminent flight first --- the passenger might of course fly more than once on the same passport, hence a number of flights would be associated to the same id, though with different dates (or times if we admit that a passenger can take off more than once in a day from the same airport). Also, passengers often use a boarding pass as a memo of the flight to take and of the gate to target, but a dedicated airport app would easily deliver the same, and more easily updated, information, also reducing the inherent risk of missing a flight. With smartphones increasingly, if not fully, pervading our lives, checking an app is bound to become better received than carrying a specific document such as a boarding card and at the same time checking the airport displays. On the other hand, although the attendant must still check that the passenger is at the right gate, her activities are dramatically reduced. The main outcome therefore is a more invisible security ceremony for both passenger and gate attendant.
This supports the claim that ceremony Flight Boarding II achieves stronger passenger authentication and authorisation than ceremony Flight Boarding I due to the reduced risk of passenger or attendant's errors, deliberate deviations or simplifications.

\begin{figure*}[h!]
\begin{center}
\begin{msc}{ceremony Flight Boarding III (face to face)}
\setlength{\instdist}{2,1cm}
\setlength{\envinstdist}{2.8cm}
\setlength{\bottomfootdist}{.3cm}
\declinst{p}{passenger}{$\mathit{P}$}
\declinst{a}{attendant}{$\mathit{A}$}
\declinst{g}{gate}{$\mathit{G}$}
\declinst{d}{airport database}{$\mathit{DB}$}
\mess{1. $\mathit{bio}$-$\mathit{info,e}$-$\mathit{id}$}{p}{g}
\nextlevel[1]
\action*{\parbox[l]{5.4cm}{SCAN $\mathit{bio}$-$\mathit{info,e}$-$\mathit{id}$\\[.5ex] VALIDATE $e$-$\mathit{id}$\\[.5ex] MATCH $\mathit{scan(bio}$-$\mathit{info), bio\_info}(e$-$\mathit{id)}$}}{g}
\nextlevel[4.5]
\mess{2. $\mathit{number(e}$-$\mathit{id})$}{g}{d}
\nextlevel
\action*{\parbox[l]{5.2cm}{QUERY $\mathit{number(e}$-$\mathit{id), current\_time}$}}{d}
\nextlevel[3]
\mess{3. $\mathit{flight\& passenger(number(e}$-$\mathit{id), current\_time})$}{d}{g}
\nextlevel
\action*{\parbox[l]{67mm}{MATCH $\mathit{flight\_number}(flight\& passenger(\cdot,\cdot))$,\\ \hspace*{12,2mm} $\mathit{flight\_number}(G, current\_time)$}}{g}
\nextlevel[3.5]
\mess{4. $e$-$\mathit{id}$}{g}{p}
\nextlevel[1.5]
\action*{\parbox[l]{6.8cm}{SET $\mathit{verdict},$\\[.5ex]if $\mathit{success(both\_activities)}$ then {\scriptsize `OPEN'} else {\scriptsize `FAIL'}}}{g}
\nextlevel[4]
\mess{5. $\mathit{verdict}$}{g}{p}
\end{msc}
\end{center}
\caption{A flight boarding ceremony where security is more invisible than in Flight Boarding II}\label{boardingIII}
\end{figure*}

I sketched this ceremony in 2016 \cite{spw16} and, borrowing systems already in use at Border Control at least in the UK, conjectured a further amendment during my talk, that ``\emph{the passport has a scan of your fingerprint, and so you just go through even without a human attendant there, you just go through, scan your passport, scan your finger, and you would be let in or not. Doesn't it work like that to enter the UK if you have an electronic passport nowadays?}'' \cite{spw16transcript}. I was speculating on a flight boarding ceremony whose security is yet more invisible for the attendant. It is ceremony ``Flight Boarding III'', here shown in Figure \ref{boardingIII}, remarkably with an \emph{empty} attendant's role. Its gist is that the electronic id that an approaching passenger hands out for the gate to scan also stores securely some biometric information, such as fingerprint or face scan, of the passenger's, which the gate matches to the homologous information scanned live from the passenger. The final check that the passenger is at the right gate is now shifted from the attendant to the gate, and the same applies to the definition of the verdict.

But spring 2016 seemed early to revolutionise the airport experience according to ceremony Flight Boarding III, and I was criticised that my idea would take too long to work at a gate, up to ten or fifteen seconds per passenger. I replied ``\emph{Yes, but tomorrow it will take three seconds. The question is, would a human attendant take less than ten seconds?}'' \cite{spw16transcript}. This has almost become reality today, with a similar ceremony ``clearing up to 10 passengers a minute'' since November 2017 through a pilot in Miami airport \cite{bioairportpdftoto}. Also British Airways has started testing similar techniques at Heathrow Terminal 5 \cite{babio}, and Dubai International airport is testing a face-scanning \emph{fish tunnel} to catch passengers' attention and make their face scanning a yet more invisible security measure \cite{fish}. These ceremonies are different but share the gist of making passenger's authentication and authorisation measures as invisible as possible, ultimately without a human attendant. The passenger's electronic id may have to be scanned just once through the entire airport experience, if at all. Even more recently, a system is being tested that leverages a database of passengers' electronic passports built over time while passengers apply for a new passport. As a result, a passenger would no longer need to carry his electronic id \cite{dubaiface} at boarding time, because his face, or any other biometric info, would be scanned and matched to the information stored in the database about the passenger. The resulting ceremony, say ``Flight Boarding IV'', omitted here, would become even leaner than Flight Boarding III, by dismissing the $e$-$\mathit{id}$ entirely.

These observation reassure us that the invisible security revolution has already began also on our planet.
With biometric scanning techniques getting more and more performing, the invisibility of security ceremonies will thrive. Earth seems to be evolving as advocated by the ``Minority Report'' film \cite{minorityreport} but at least the privacy implications are yet to be fully explored. \rev{However, removing the human component from the enforcement of security measures may also have negative consequences that are yet to be explored fully. For example, a human attendant may observe passengers's suspicious behaviour and notice countless elements that technology would not unless it is specifically programmed to do so}.

\section{Discussion}\label{disc}

\begin{table*}[h]
\begin{center}
\begin{tabular}{l||c|c|c|c|c|c||}
	&\multicolumn{2}{|c|}{Dictatorial} & \multicolumn{2}{|c|}{Beautiful} & \multicolumn{2}{|c|}{Invisible}\\ 
	\hline
	\hline  	
	 Unconstrai\-ned Password Setup (\S\ref{dict})& \multicolumn{2}{|c|}{\xmark} &\multicolumn{2}{|c|}{\checkmark?} & \multicolumn{2}{|c||}{\xmark}\\
	 \hline
	 NIST 2004 Password Setup (\S\ref{dict})& \multicolumn{2}{|c|}{$\sim$} &\multicolumn{2}{|c|}{\xmark?} & \multicolumn{2}{|c||}{\xmark}\\
	 \hline
	 NIST 2017 Password Setup (\S\ref{dict})& \multicolumn{2}{|c|}{\checkmark} &\multicolumn{2}{|c|}{\checkmark?} & \multicolumn{2}{|c||}{\xmark}\\
	 \hline
	 Gamified Password Setup (\S\ref{beauty})& \multicolumn{2}{|c|}{\checkmark} &\multicolumn{2}{|c|}{\checkmark} & \multicolumn{2}{|c||}{\xmark}\\
	 \hline\hline
	 Current WhatsApp Web Login (\S\ref{beauty})& \multicolumn{2}{|c|}{\checkmark} &\multicolumn{2}{|c|}{\checkmark} & \multicolumn{2}{|c||}{\xmark}\\
	 \hline
	 Tap-in WhatsApp Web Login (\S\ref{beauty})& \multicolumn{2}{|c|}{\checkmark} &\multicolumn{2}{|c|}{\xmark} & \multicolumn{2}{|c||}{\xmark}\\
	\hline 
	\hline
iPhone 5 Wake-up (\S\ref{disappearance})& \multicolumn{2}{|c|}{\xmark} &\multicolumn{2}{|c|}{\xmark?} & \multicolumn{2}{|c||}{\xmark}\\
	 \hline
iPhone 5S Wake-up (\S\ref{disappearance})& \multicolumn{2}{|c|}{\xmark} &\multicolumn{2}{|c|}{\checkmark?} & \multicolumn{2}{|c||}{\checkmark}\\
	 \hline
	 \hline
&passenger&attendant&passenger&attendant&passenger&attendant\\
\hline
	 Flight Boarding I (\S\ref{disappearance})& \checkmark & \xmark& \xmark& \xmark& \xmark&\xmark\\
	 \hline
	 Flight Boarding II (\S\ref{disappearance})& $\sim$& \xmark & $\sim$ &\checkmark? &$\sim$ &$\sim$\\
	 \hline
	 Flight Boarding III (\S\ref{disappearance})& $\sim$& $\varnothing$ & $\sim$ & $\varnothing$ & $\sim$&\checkmark\\
	 \hline
	 Flight Boarding IV (\S\ref{disappearance})& \checkmark& $\varnothing$ & \checkmark & $\varnothing$ &\checkmark &\checkmark\\
	 \hline
	 \hline
\end{tabular} 
\end{center}
\caption{Security features of relevant example ceremonies}\label{table:comparison}
\end{table*}

All cities are distinct, and it is clear that only the 
ceremonies that are used in Democratic City ignore the human factor; hence, such ceremonies cannot be used in other cities, whose ceremonies, conversely, attempt to address that factor.

Therefore, a ceremony may exist in more than one city among the other three,as summarised in Table \ref{table:comparison}, which is discussed below. I also conjecture that ceremonies may exist outside cities, combining partial features of the neighbouring cities. 

Table \ref{table:comparison} compares the features of being dictatorial, beautiful or invisible that the most relevant ceremonies introduced above exhibit. 
To offer a useful summary, contents are simplified to tick, cross and tilde, with the latter indicating \emph{some} existence of a feature. 
Defining the precise range of each feature exceeds the scope of this article, therefore quantitative arguments on the extent to which a feature is present, or on which of two ceremonies has more of a feature cannot be made here. 
Likewise, one may be left wondering whether the Current WhatsApp Web Login ceremony could be made yet more beautiful or whether there exists a flight boarding ceremony resulting yet
less dictatorial for the attendant than Flight Boarding I (arguably the former question seems more relevant than the latter). However, the top level of invisibility for a participant is easy to define through the emptying of the participant's role, as with the attendant's in Flight Boarding III. The empty set symbol stands for the empty role, and the question mark is often used on the `beautiful' feature to indicate the inherent subjectivity.

If a security ceremony invites more than a single human role, then the three features may vary across participants and, therefore, must be evaluated on each. This is the case with flight boarding, at least originally.  
So, every column of Table \ref{table:comparison} forks on each flight boarding ceremony. 

We know that Dictatorial City cancels human choices during the interaction with the ceremonies. Therefore, it is clear that Unconstrained Password Setup is not dictatorial. Equal\-ly, the first two boarding ceremonies are not dictatorial for the attendant and, in particular, do not prescribe anything like having the attendant fulfill an activity checklist that is guard\-ed and guaranteed by someone else or by machines. Both iPhone wake-up ceremonies are not dictatorial because the human can decide to remove their security measure and wake up a phone without any authentication at all.

It can be seen that six other ceremonies are dictatorial, including Flight Boarding IV for the passenger, because the passenger cannot influence the security of the measure (because it is invisible). Three other ceremonies are somewhat dictatorial, as the tilde indicates. As mentioned (\S\ref{beauty}), the older version of the NIST password setup ceremony turns out less dictatorial than the newer version because only the older allows the human to choose the password. It could be further investigated whether Flight Boarding III is received as more dictatorial than Flight Boarding II by random passengers. 
Whether to consider the last two boarding ceremonies dictatorial for the attendant despite her empty role seems a philosophical question, hence I prefer to merely indicate the empty role.

Compliance of the ceremonies to the beautiful security principle should be broadly explored in general. This explains the question marks next to my best evaluations: positive for Unconstrained Password Setup due to a sense of freedom, for NIST 2017 Password Setup due to a sense of simplicity, and for Iphone 5S Wake-up due to a sense of fun; negative for NIST 2004 Password Setup and for Iphone 5 Wake-up due to senses of frustration and boredom. Furthermore, I postulate that the Gamified Password Setup ceremony conforms to the principle, as well as the Current WhatsApp Web Login ceremony does, in contrast to its tap-in version. Likewise, I postulate that Flight Boarding IV is beautiful for the passenger.

The table also shows that Flight Boarding I is not beautiful for either participants, and that versions II and III are somewhat beautiful for the passenger (hence more beautiful than version I is), although deciding which of II or III is more beautiful needs a fuller evaluation. I also argue that version II is beautiful for the attendant, while her role gets emptied in the subsequent ceremonies. It can be concluded overall that evolving from version I onward increases compliance to beautiful security.

A similar conclusion can be drawn about the flight boarding ceremonies in terms of invisible security, as it can be gathered from the last column. 
The evaluation of this feature for the flight boarding passenger is the same as that of beautiful security.
Then, version II is more invisible than version I for the attendant, and both versions III and IV are invisible for her due to an empty role.
The higher part of the column is easy to interpret, with only the iPhone 5S Wake-up ceremony being invisible.

\section{Conclusions}\label{concl}
Cybersecurity is a planet with lots and lots yet to be discovered. I have only spotted four cities on it thus far, and most of what I argue about them derives from projections of terrestrial experience. 

Democratic City overlooks the inherent socio-technical aspects of the \s{} problem, although the problem relates to both the technical features of the \s{} ceremonies and to how humans receive and avail themselves of such ceremonies. Because humans may make errors, deliberately oppose the \s{} ceremonies or be deceived, I have argued that common vulnerabilities and related exploitations in this city are \emph{also} due to an insufficient account on the human factor during the design of the ceremonies.

All other cities share the feature of addressing the socio-technical aspects of the \s{} problem, though each in a different way, namely reducing, if not removing, the potential risk that security ceremonies falter due to the way humans (ab)use them. Dictatorial City sees ceremonies that are fully prescriptive on humans, thereby removing the risks deriving from human interaction. Security ceremonies in Beautiful City achieve the same goal by being beautiful, hence by attracting humans to comply. Invisible City is perhaps the most advanced, combining measures that are there but work without posing any apparent demand on humans.

The discovery and exploration of the \Cs{} planet have been very illuminating so far, helping distill ideas and evolve the \s{} ceremonies to address the human factor of the security problem. This article supports the claim that dictatorial security, beautiful security and invisible security are three distinct approaches to strengthen security ceremonies \rev{that currently implement democratic security. The resulting \s{} ceremonies will be} effective no matter the interaction with humans.
Which approach, or  combination of them, is best for large-scale deployment in general is not fully understood yet; it might be the case that \rev{it will be depending on the target application scenario that} each \s{} measure (and adopting ceremony) will find its best-suited approach to strengthening it for widespread human use. \rev{Therefore, we need to study how to decide, first of all, the best approach, whether dictatorial, beautiful or invisible \s, to improve a given security ceremony. Then, we need to increase our efforts to research and develop concrete, general methodologies to take each approach. As anticipated above, such efforts cannot be anything but inter-disciplinary.}

\rev{I conclude conjecturing that the \Cs{} planet may still hide further approaches to improving \s{} measures and establishing \s{} in practice.} 
We do not know \rev{very} much about that far planet, but what can be  surmised convinces me that its continuous exploration will surprise and enrich us \rev{a great deal}.

\Urlmuskip=0mu plus 1mu\relax
\bibliographystyle{plain}
\bibliography{biblio}

\begin{thebibliography}{10}

\bibitem{babio}
Mike Arnot.
\newblock {Your Face Is The Token: We Tried British Airways’ New Biometric
  Screening}, 2018.
\newblock
  \url{https://thepointsguy.com/2018/01/we-tried-british-airways-new-biometric-screening/}.

\bibitem{jbrowser}
G.~Bella, R.Giustolisi, and G.Lenzini.
\newblock {Invalid Certificates in Modern Browsers: A Socio-Technical
  Analysis}.
\newblock {\em {IOS} Journal of Computer Security}, 26(4):509--541, 2018.

\bibitem{spw16}
Giampaolo Bella, Bruce Christianson, and Luca Vigan{\`o}.
\newblock {Invisible Security}.
\newblock In Jonathan~Anderson et~al., editor, {\em Proceedings of the 24th
  International Workshop on Security Protocols (SPW'16)}, LNCS 10368, pages
  1--9. Springer, March 2016.

\bibitem{spw16transcript}
Giampaolo Bella, Bruce Christianson, and Luca Vigan{\`o}.
\newblock {Invisible Security (Transcript of Discussion)}.
\newblock In Jonathan~Anderson et~al., editor, {\em Proceedings of the 24th
  International Workshop on Security Protocols (SPW'16)}, LNCS 10368, pages
  1--9. Springer, March 2016.

\bibitem{spw15}
Giampaolo Bella and Luca Vigan{\`o}.
\newblock {Security is Beautiful}.
\newblock In {\em Proceedings of the 23rd International Workshop on Security
  Protocols (SPW'15)}, LNCS 9379, pages 247--250. Springer, 2015.

\bibitem{nist}
William~E Burr, Donna~F Dodson, and W~Timothy Polk.
\newblock Nist special publication 800-63, 2004.

\bibitem{disappearing}
Josiah Dykstra and Eugene Spafford.
\newblock The case for disappearing cyber security.
\newblock {\em Communications of the ACM}, 61, 07 2018.

\bibitem{fish}
Annabel~Fenwick Elliott.
\newblock {Dubai Airport is replacing security checks with face-scanning fish},
  2017.
\newblock
  \url{https://www.telegraph.co.uk/travel/news/dubai-airport-replaces-security-checks-with-face-scanning-fish/}.

\bibitem{Ellison2007}
Carl Ellison.
\newblock Ceremony design and analysis, 2007.
\newblock {IACR eprint archive}.

\bibitem{tls-ceremony}
R.~Giustolisi, G.~Bella, and G.Lenzini.
\newblock {Invalid Certificates in Modern Browsers: A Socio-Technical
  Analysis}.
\newblock {\em {IOS Journal of Computer Security}}, 26(4):509--541, 2018.

\bibitem{nistnewreport}
Paul~A. Grassi, James~L. Fenton, Elaine~M. Newton, Ray~A. Perlner, Andrew~R.
  Regenscheid, William~E. Burr, and Justin~P. Richer.
\newblock Nist special publication 800-63, 2017.
\newblock \url{https://pages.nist.gov/800-63-3/sp800-63b.html}.

\bibitem{2018pwd}
John Hall.
\newblock {SplashData's Top 100 Worst Passwords of 2018}, 2018.
\newblock
  \url{https://www.teamsid.com/splashdatas-top-100-worst-passwords-of-2018/}.

\bibitem{hendrix2016game}
Maurice Hendrix, Ali Al-Sherbaz, and Bloom Victoria.
\newblock Game based cyber security training: are serious games suitable for
  cyber security training?
\newblock {\em International Journal of Serious Games}, 3(1):53--61, 2016.

\bibitem{karenic}
Rossouw Von~Solms Karen~Renaud, Basie Von~Solms.
\newblock {How does intellectual capital align with cyber security?}
\newblock {\em {Journal of Intellectual Capital}}, 2019.

\bibitem{gamauth}
C.~Kroeze and M.~S. Olivier.
\newblock {Gamifying Authentication}.
\newblock In {\em {Proceedings of Information Security for South Africa
  (ISSA'12)}}. IEEE Press, 2012.

\bibitem{channels}
Jean~Everson Martina, Eduardo dos Santos, Marcelo~Carlomagno Carlos, Geraint
  Price, and Ricardo~Felipe Cust\'odio.
\newblock {An adaptive threat model for security ceremonies}.
\newblock {\em {International Journal of Information Security}}, 14:103--121,
  2014.

\bibitem{mitnick}
Kevin~D. Mitnick and William~L. Simon.
\newblock {\em {The Art of Deception: Controlling the Human Element of
  Security}}.
\newblock John Wiley \& Sons, 2001.

\bibitem{norman}
Donald~A. Norman.
\newblock Categorization of action slips.
\newblock {\em Psychological Review}, 88(1):1--15, 1981.

\bibitem{psychoschneier}
Bruce Schneier.
\newblock The psychology of security.
\newblock In {\em Proceedings of the Cryptology in Africa 1st International
  Conference on Progress in Cryptology (AFRICACRYPT'08)}, pages 50--79.
  Springer, 2008.

\bibitem{shellshock}
{Symantec Security Response}.
\newblock {ShellShock: All you need to know about the Bash Bug vulnerability},
  2014.
\newblock
  \url{https://www.symantec.com/connect/blogs/shellshock-all-you-need-know-about-bash-bug-vulnerability}.

\bibitem{heartbleed}
{United States Computer Emergency Readiness Team}.
\newblock {OpenSSL 'Heartbleed' vulnerability (CVE-2014-0160)}, 2014.
\newblock \url{https://www.us-cert.gov/ncas/alerts/TA14-098A}.

\bibitem{bioairportpdftoto}
Unknown.
\newblock {Biometrics land in more airports worldwide}.
\newblock {\em Biometric Technology Today}, 2018(3):12, 2018.
\newblock \url{https://books.google.it/books?id=ikHfvQEACAAJ}.

\bibitem{peppa}
URL.
\newblock {Peppa Pig, Series 3, Episode 38, ``The Secret Club''}, 2010.
\newblock \url{https://www.youtube.com/watch?v=QSQhScDvOao}.

\bibitem{ryanairpisa}
URL.
\newblock {Ryanair passenger gets on wrong plane and flies to Sweden instead of
  France}.
\newblock
  \url{https://www.independent.co.uk/travel/news-and-advice/ryanair-passenger-wrong-place-flight-bari-sardinia-cagliari-video-a8926921.html},
  2012.

\bibitem{Mirror}
URL.
\newblock {Ryanair passenger lands in wrong Italian city}, 2012.
\newblock
  \url{http://www.mirror.co.uk/news/uk-news/ryanair-passenger-gets-on-wrong-plane-946207}.

\bibitem{boxdropbox}
URL.
\newblock {Dropbox, Box users Leak Sensitive Information via Shared Links
  Flaw}, 2014.
\newblock
  \url{http://techfrag.com/2014/05/08/dropbox-box-users-leak-sensitive-information-via-shared-links-flaw/}.

\bibitem{chatham}
URL.
\newblock {Chatham House Report}.
\newblock \url{https://shorturl.at/msuEK}, 2015.

\bibitem{ibm}
URL.
\newblock {IBM Security Services 2014 Cyber Security Intelligence Index}, 2015.
\newblock
  \url{https://www.ibm.com/developerworks/library/se-cyberindex2014/index.html}.

\bibitem{fridge}
URL.
\newblock {Samsung smart fridge leaves Gmail logins open to attack}, 2015.
\newblock
  \url{http://www.theregister.co.uk/2015/08/24/smart_fridge_security_fubar/}.

\bibitem{cranor}
URL.
\newblock {Frequent password changes are the enemy of security, FTC
  technologist says}, 2016.
\newblock
  \url{http://arstechnica.com/security/2016/08/frequent-password-changes-are-the-enemy-of-security-ftc-technologist-says/}.

\bibitem{amazon}
URL.
\newblock {STEELMASTER Swipe Card or Badge Rack}, 2016.
\newblock
  \url{https://www.amazon.com/STEELMASTER-Swipe-Capacity-Inches-20401/dp/B002V85VWQ}.

\bibitem{ibm18}
URL.
\newblock {IBM X-Force Threat Intelligence Index 2018}, 2018.
\newblock
  \url{https://www-01.ibm.com/common/ssi/cgi-bin/ssialias?htmlfid=77014377USEN}.

\bibitem{pokayoke}
URL.
\newblock {Poka-yoke}, 2019.
\newblock \url{https://en.wikipedia.org/wiki/Poka-yoke}.

\bibitem{verizon19}
URL.
\newblock {Verizon 2019 Data Breach Investigations Report}, 2019.
\newblock \url{https://enterprise.verizon.com/resources/reports/dbir/}.

\bibitem{krack}
Mathy Vanhoef and Frank Piessens.
\newblock Key reinstallation attacks: Forcing nonce reuse in {WPA2}.
\newblock In {\em Proceedings of the 24th ACM Conference on Computer and
  Communications Security (CCS'17)}. ACM, 2017.

\bibitem{psychowest}
Ryan West.
\newblock The psychology of security.
\newblock {\em {Communications of the ACM}}, 51(4):34--40, April 2008.

\bibitem{minorityreport}
Wikipedia.
\newblock {Minority Report (film)}, 2002.
\newblock \url{https://en.wikipedia.org/wiki/Minority_Report_(film)}.

\bibitem{dubaiface}
Dean Wilkins.
\newblock {Enter the UAE with just your face}, 2018.
\newblock
  \url{https://www.timeoutdubai.com/dubai-airport/386146-enter-the-uae-with-just-your-face}.

\end{thebibliography}

\end{document}